\documentclass[11pt,twoside]{article}
\usepackage{./asp2010}
\usepackage{subfigure}

\resetcounters

\newcommand{\Msun}{{\ensuremath{\it{M}_{\odot}}}}
\newcommand{\Zsun}{\ensuremath{Z_\odot}}

\newcommand{\K}{\ensuremath{\mathrm{K}}}
\newcommand{\cm}{\ensuremath{\mathrm{cm}}}
\newcommand{\gcc}{\ensuremath{\mathrm{g}\,\mathrm{cm}^{-3}}}

\newcommand{\Ni}{{\ensuremath{^{56}\mathrm{Ni}}}}
\newcommand{\Fe}{{\ensuremath{^{56}\mathrm{Fe}}}}

\newcommand{\Co}{{\ensuremath{^{56}\mathrm{Co}}}}

\newcommand{\CASTRO}{\texttt{CASTRO}}

\begin{document}

\title{Pair-Instability Supernovae of Non-Zero Metallicity Stars}
\author{Ke-Jung Chen$^{1,2,*}$, Alexander Heger$^{2,3,4}$, 
Stan Woosley$^{1}$, Ann Almgren$^{5}$ and Daniel J. Whalen$^{6}$ 
\affil{$^1$Department of Astronomy \& Astrophysics, University of California, Santa 
Cruz, CA 95064, USA}
\affil{$^2$School of Physics and Astronomy, University of Minnesota, Minneapolis, MN 
55455, USA}
\affil{$^3$Monash Centre for Astrophysics, Monash University, Victoria 3800, Australia }
\affil{$^4$Joint Institute for Nuclear Astrophysics, University of Notre Dame Notre Dame, IN 46556 USA} 
\affil{$^5$Center for Computational Sciences and Engineering, Lawrence Berkeley 
National Lab, Berkeley, CA 94720, USA}
\affil{$^6$Zentrum f\"{u}r Astronomie, Institut f\"{u}r Theoretische Astrophysik, 
Universit\"{a}t Heidelberg, Albert-Ueberle-Str. 2, 69120 Heidelberg, Germany}
\affil{$^*$IAU Gruber Fellow; kchen@ucolick.org}
}

\begin{abstract}

Observational evidence suggests that some very massive stars in the local Universe may die 
as pair-instability supernovae.  We present 2D simulations of the pair-instability supernova of 
a non-zero metallicity star.  We find that very little mixing occurs in this explosion because 
metals in the stellar envelope drive strong winds that strip the hydrogen envelope from the 
star prior to death.  Consequently, a reverse shock cannot form and trigger fluid instabilities
during the supernova.  Only weak mixing driven by nuclear burning occurs in the earliest 
stages of the supernova, and it is too weak to affect the observational signatures of the
explosion.

\end{abstract}

\section{Introduction}

The fate of a massive star is determined by its initial mass, composition, and mass loss over
its life (which is still poorly understood).  The explosion mechanism and remnant properties 
is mainly determined by the helium core mass when the star dies. \citet{kudri2002} suggests 
that the mass loss rate of a star, $\dot{m}$, is $\propto\,Z^{0.5}$, where $Z$ is the metallicity 
of the star relative to the solar metallicity, $\Zsun$.  Since the first (or Pop~III) stars form in 
pristine H and He gas, it is generally thought that they retain most of their mass over their 
lifetimes.  Pop~III stars above $80\,\Msun$ encounter the pair production instability after 
central carbon burning, in which large numbers of thermal photons create $e^-/e^+$ pairs at 
the expense of pressure support in the core.  If Pop~III stars are over $150\,\Msun$ but less 
than $260\,\Msun$, core contraction due to the loss of pressure support to pair production 
ignites explosive oxygen and silicon burning that results in an energetic explosion that 
completely unbinds the star.  This thermonuclear explosion is called a pair-instability 
supernova (PSN; \citealt{barkat1967, heger2002, chen2014r, chen2014psn, chen2014ppsn}).  
A PSN can explode with up to $10^{53}$ erg of energy, or about 100 times that of a Type Ia SN.  
Explosive silicon burning can synthesize up to 40 \Msun\ of \Ni{} in these events which, along 
with their large explosion energies, makes them extremely bright.  PSNe are therefore visible 
at high redshifts and could be used to probe the properties of the first stars \citep{kasen2011,
wet12a,wet12b,pan12a}. Isotopes heavier than the iron group are completely absent from the 
chemical yields of PSNe because of the absence of neutron capture processes (r- and 
s-processes).  The important question arises : Can the most massive stars in the local Universe 
also die as PSNe?  If so, how do these PSNe differ from those of Pop~III progenitors?

The detection of PSN candidates SN 2007bi and SN 2213 - 1745 \citep{galyam2009,cooke12} 
has increased interest in PSNe and poses a challenge to theories of galactic star formation, 
because how progenitors so massive can form at metallicities of $0.1\,\Zsun$ today is very 
unclear. Stars at these metallicities can easily lose most of their mass over their lifetimes to 
strong stellar winds, and so they must form at even higher masses than Pop III stars to die 
as PSNe.  Understanding the observational signatures of PSNe at near-solar metallicities is 
key to identifying and properly interpreting these events as more are discovered.  To date, 
PSN models have focused on Pop III stars.  In this paper, we present 2D simulations of a PSN 
of a non-zero metallicity star and investigate if mixing can affect explosive burning or 
observational signatures.  We first describe our numerical methods and problem setup in 
\S~\ref{solpsn_Methodology}.  Our results in are discussed in \S~\ref{solpsn_results}, and our 
conclusions are given in \S~\ref{solpsn_conclusions}.

\section{Methodology \& Problem Setup}

\label{solpsn_Methodology}

Self-consistent multidimensional stellar evolution models from the onset of hydrogen burning to 
eventual core collapse and explosion remain beyond the realm of contemporary computational 
power.  We instead initialize our multidimensional explosion simulation with a $500\,\Msun$ 0.1
\Zsun\ star that is evolved from the zero-age main sequence to the onset of core collapse in the
1D stellar evolution code {\texttt{GENEVA}} \citep{hirschi2004, whalen2013}. Our model includes 
mass loss and stellar rotation to determine the final structure of the star.  The initial rotational 
velocity is $40\,\%$ of the critical velocity, or a surface velocity of 450 km/sec at the equator. 
The star is evolved up to the onset of explosive oxygen burning, about 10 s before maximum 
core contraction.  Initializing this profile in a 2D simulation at this time should therefore capture
instabilities seeded by both collapse and explosive burning.  The explosion time scale is about 
tens of seconds which is much shorter than the rotational period of star, $P\sim$ several hours. 
It is reasonable to assume that the star is non-rotating during the explosion phase, so we neglect 
the effect of rotation in the 2D simulation. 
 
We map the 1D profile of the star to a 2D cylindrical coordinate grid in $r$ and $z$ in \CASTRO{} 
with the conservative mapping scheme of \citet{chen2013}.  \CASTRO{} is a multidimensional 
adaptive mesh refinement (AMR) astrophysical radiation hydrodynamics code \citep{ann2010,
zhang2011}.  It has a higher-order unsplit Godunov hydro scheme and block-structured AMR.
In our models we use the Helmholtz EOS \citep{timmes2000} for stellar matter, which includes 
degenerate and non-degenerate, relativistic and non-relativistic electrons, electron-positron pair 
production, and an ideal gas with radiation.  \CASTRO{} evolves mass fractions for each isotope 
with its own advection equation. The monopole approximation is used for self-gravity, in which a 
1D gravitational potential is constructed from the radial average of the density and then 
differenced to construct the gravitational force vector everywhere in the AMR hierarchy.  This 
approximation is well-suited to the nearly spherical symmetry of the star and is efficient. We use 
the 19-isotope APPROX nuclear reaction network \citep{kepler,timmes1999}, which includes 
heavy-ion reactions, alpha-chain reactions, hydrogen burning cycles, photo-disintegration of 
heavy nuclei, and energy loss through thermal neutrinos.  Nuclear burning is self-consistently
coupled to hydrodynamics, and we also account for energy deposition due to radioactive decay 
of \Ni{} $\to$ \Co{} $\to$ \Fe{}. 

The grid size is $4\cdot10^{10}\times4\cdot10^{10}\,\cm^2$, with 256 uniform zones in both $r$
and $z$. Up to three levels of refinement, each with a factor of 4 greater resolution, are permitted
throughout the simulation.  The resolution at the finest level is $2\times10^7\,\cm$, which is 
sufficient to resolve the characteristic scales of nuclear burning ($\sim 10^8\,\cm$).  The grid is
refined on functions of the gradients in density, velocity, and pressure.  Because we only simulate 
one octant of the star, we apply reflecting boundary conditions at the inner $r$ and $z$ boundaries;
we impose outflow boundary conditions on the upper boundaries.  Refined grids are nested around 
the core of the star to assure that it is always at the highest resolution.   We evolve our simulations 
until the shock breaks through the surface of the star. 

\begin{figure}[h]
\vspace{-.2cm}
\begin{center} 
\subfigure[Mass fraction of selected isotopes]{\label{x500}\includegraphics[width=0.46\textwidth]{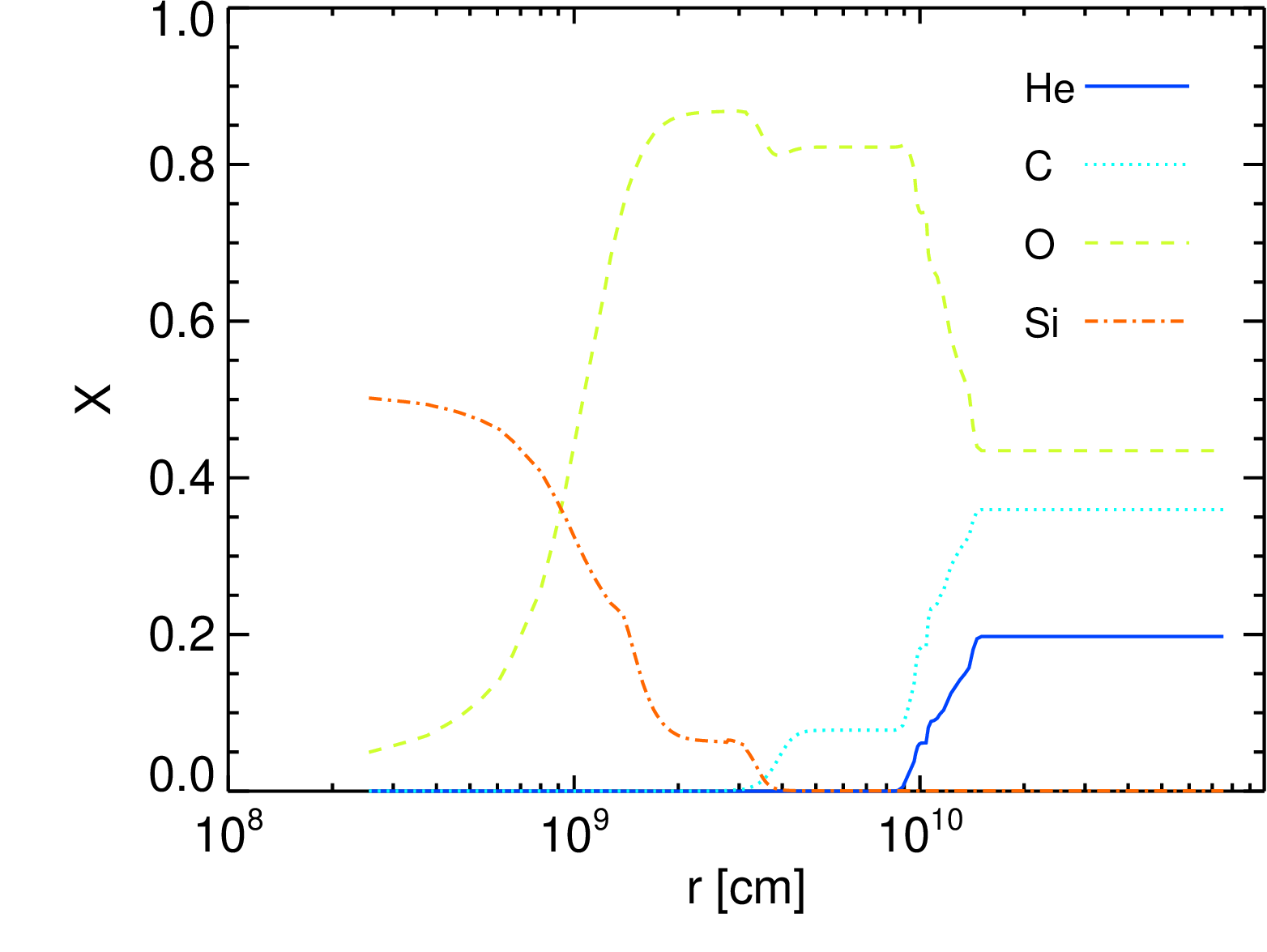}}   
\subfigure[$\rho r^3$ plot]{\label{p500_rhor3}\includegraphics[width=0.46\textwidth]{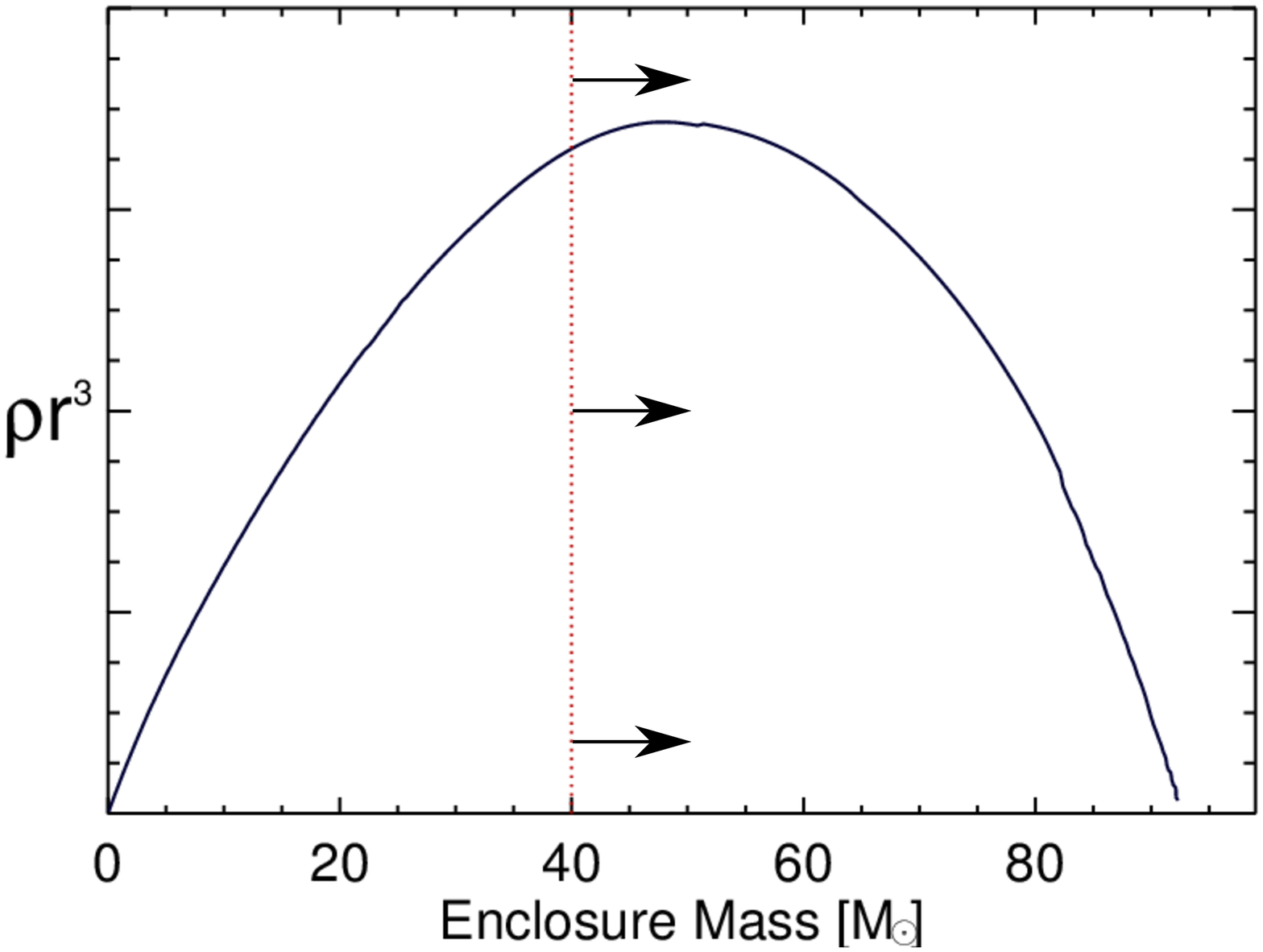}}                

\caption{Left:  isotope mass fractions. The high silicon mass fraction at the core shows that oxygen 
burning is complete.  Right:  plot of $\rho r^3$.  The bump at the center is due to the helium core.  
The red dashed line marks the approximate site of the formation of the shock.  The shock 
propagates only a short distance through a region of increasing $\rho r^3$, so it is unlikely to 
decelerate and create a reverse shock or therefore fluid instabilities.}
\end{center}
\end{figure}

\section{Results}

\label{solpsn_results}

\begin{figure}[h]
\vspace{-.2cm}
\begin{center} 
\subfigure[Fluid instabilities during the explosion]{\label{2d_sol_burning} 
\includegraphics[width=0.48\textwidth]{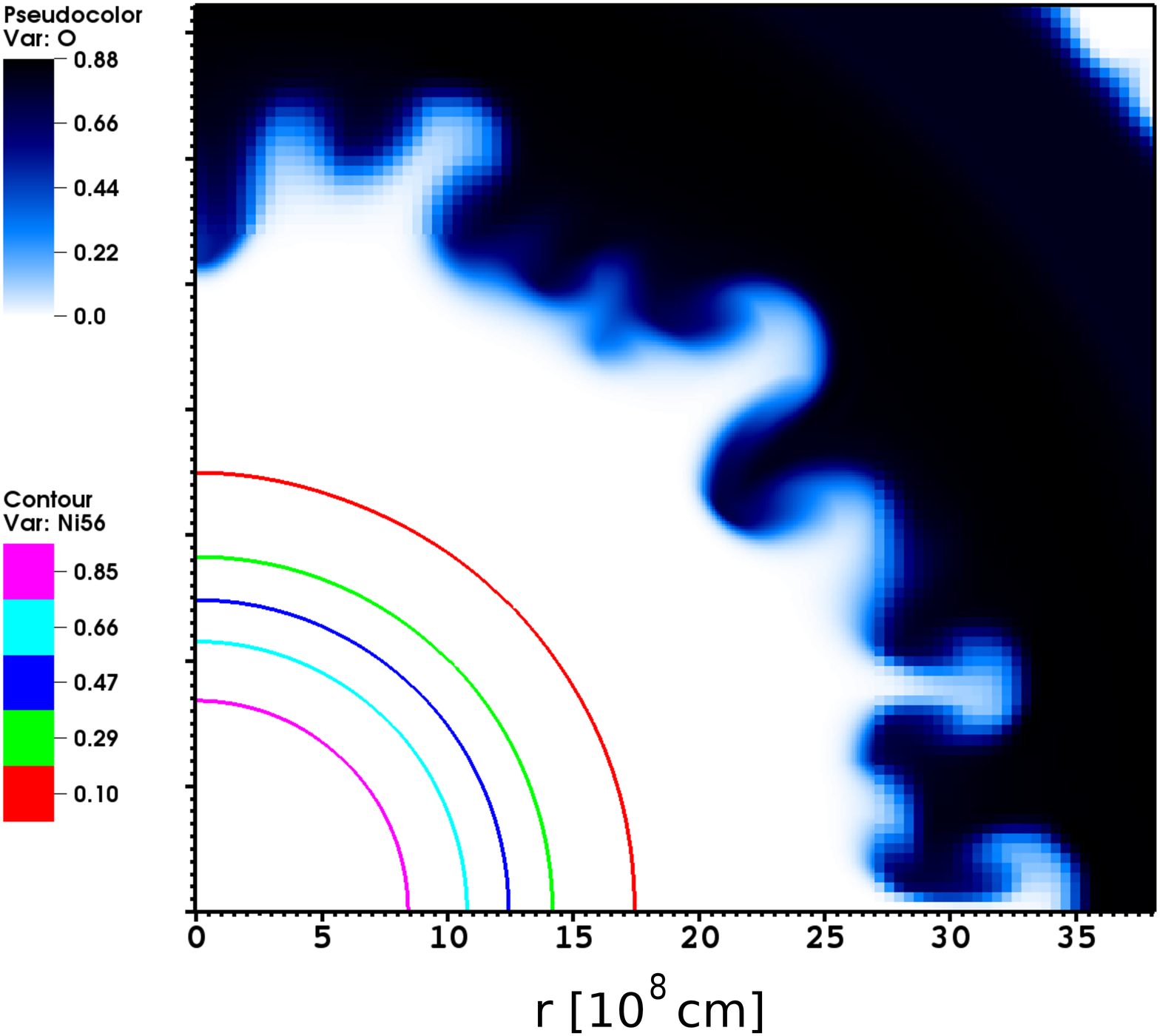}}   
\subfigure[Shock breakout]{\label{2d_sol_breakout}
\includegraphics[width=0.46\textwidth]{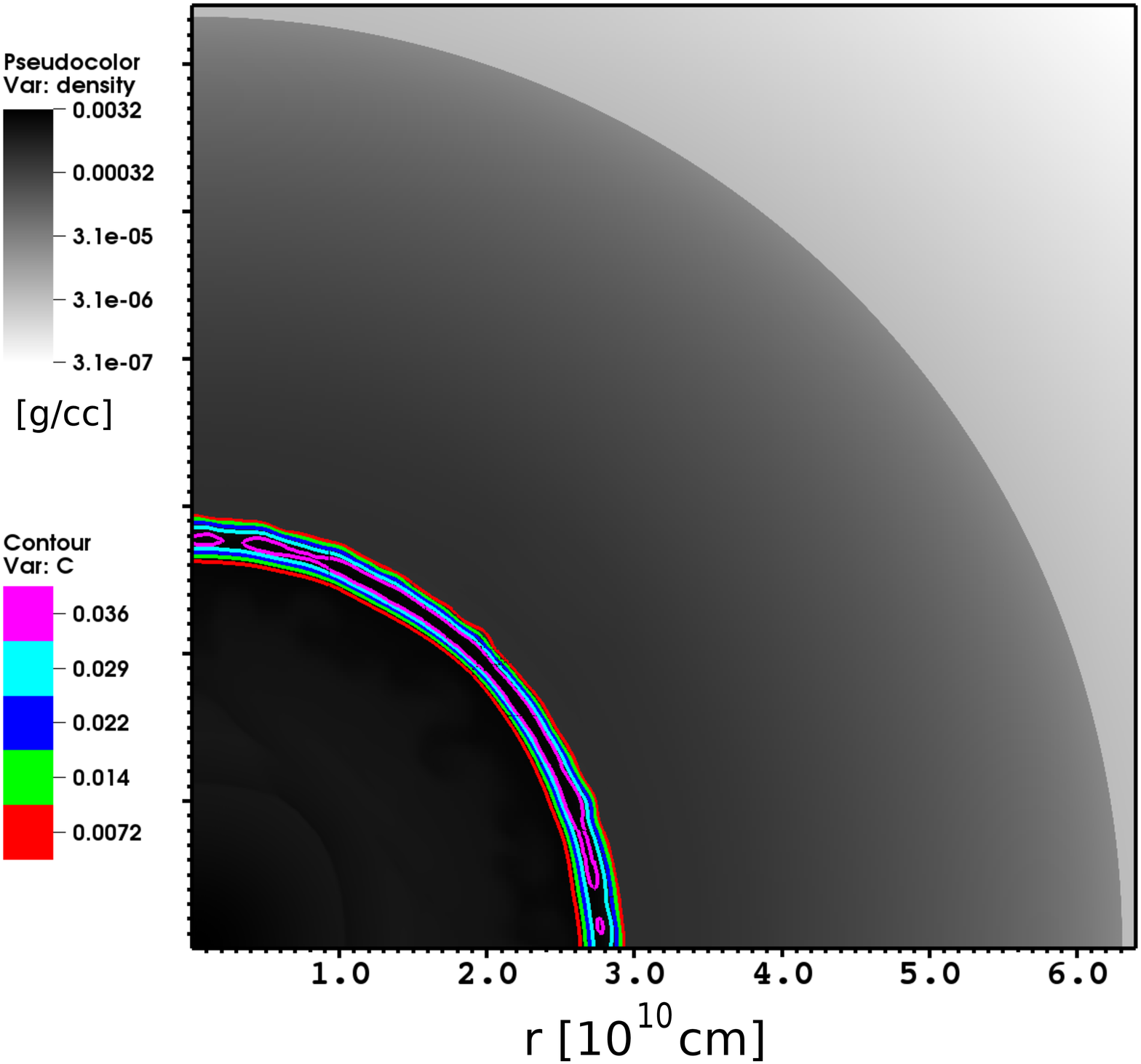}}                
\caption{Left:  oxygen and \Ni\ mass fractions.  \Ni\ mass fractions within the red arc are greater than $0.1$. 
The inner region of the oxygen-burning shell shows only mild mixing during explosive burning that
does not reach the \Ni\ layer.  Right:  densities (gray) and carbon abundance (contours). 
 The SN ejecta is fairly spherical, and there is no evidence of mixing in density or mass fractions of carbon.}
\end{center}
\end{figure}

The mass of the star falls from $500\,\Msun$ to $92.5\,\Msun$ over its lifetime because of stellar 
winds and rotation.  Strong winds not only remove the hydrogen envelope but also strip away the 
outer layer of the helium core. The central temperature and density of the core when it has begun 
to contract are $3.31\times10^9\,\K$ and $1.24\times10^{6}\,\gcc$. In Figure~\ref{x500} we show 
mass fractions for helium, carbon, oxygen, and silicon at the end of the GENEVA simulation.  The 
oxygen is nearly depleted, and explosive silicon burning is about to begin because of runaway core 
contraction.  Explosive silicon burning begins soon after the \CASTRO{} run is launched, after a 
brief phase of further contraction.  Burning releases $3.33\times 10^{52}$ ergs and synthesizes 
$3.63\,\Msun$ of \Ni.  The star is completely disrupted, leaving no compact object behind.  This 
large \Ni{} mass can power the PSN light curve for several months.  If fluid instabilities dredge \Ni\
up from greater depths, it could affect luminosities at intermediate times. We plot oxygen and \Ni\ 
mass fractions $10$ s after core bounce in Figure~\ref{2d_sol_burning}. Moderate fluid instabilities 
driven by nuclear burning are visible at the inner boundary of the oxygen shell, but mixing is limited. 
The \Ni{} essentially remains untouched by dynamical instabilities.

The explosion drives a strong shock with initial velocities of $2-3\times10^9$ cm/s.  Because the 
radius of the star is only $\sim 3\times 10^{10}$ cm, the shock reaches its surface in under $20$s. 
Breakout happens at a much smaller radius than in Pop III PSNe (10$^{12}$ - 10$^{13}$ cm). We 
show densities and oxygen abundances at shock breakout in 
Figure~\ref{2d_sol_breakout}.  The SN ejecta remains roughly spherical in density and O mass 
fraction, indicating there is not much mixing during the explosion.   There are no prominent signs 
of mixing during collapse, burning 
or subsequent expansion.  The reason for this can be found in the plot of $\rho r^3$ in the right
panel of Figure 1.  The shock decelerates in regions of increasing $\rho r^3$ as it plows up mass,
and could form a reverse shock that is prone to Rayleigh-Taylor (RT) instabilities.  But Figure~\ref{p500_rhor3}
shows that when the SN shock forms it is nearly out of the region of increasing $\rho r^3$, so there
is little time for RT instabilities, or mixing, to occur.

\section{Conclusions}
\label{solpsn_conclusions}

We have examined mixing in the PSN explosions of compact non-zero metallicity stars.  We 
find that only mild fluid instabilities form during the explosion and do not result in visible mixing 
at shock breakout. This is primarily due to the fact that RT instabilities do not have time to form 
before the shock breaks out of the compact star, which is essentially a bare He core.  In this 
sense, such explosions may be similar to Pop~III PSNe of blue progenitors.  Our model has an 
explosion energy and \Ni{} yield that is similar to the $200\,\Msun$ Pop~III PSN in \citet{
chen2014psn}.  

The formation of a $500\,\Msun$ star at near-solar metallicities is a challenge to current theories 
of present-day star formation.  Radiation from the star normally halts accretion well before the 
star reaches such masses.  Mergers between stars in dense clusters in principle could build up 
stars of up to several hundred solar masses.  If so, dense star clusters could be promising sites 
for hunting for pair-instability supernovae in the local Universe.

\acknowledgements KC thanks the members of the CCSE at LBNL for help with \CASTRO{}. 
This work was supported by the IAU-Gruber Fellowship, Stanwood Johnston Fellowship, and 
KITP Graduate Fellowship. Work at UCSC has been supported by the DOE HEP Program 
under contract DE-SC0010676; the National Science Foundation (AST 0909129) and the
NASA Theory Program (NNX09AK36G).  AH acknowledges support by an ARC Future Fellowship (FT120100363) and a 
Monash University Larkins Fellowship.  DJW acknowledges support from the European Research Council 
under the European Community's Seventh Framework Programme (FP7/2007-2013) via the 
ERC Advanced Grant "STARLIGHT: Formation of the First Stars" (project number 339177).  
All numerical simulations were done with allocations from the University of Minnesota 
Supercomputing Institute and the National Energy Research Scientific Computing Center.  


\begin{thebibliography}{}
\expandafter\ifx\csname natexlab\endcsname\relax\def\natexlab#1{#1}\fi
\expandafter\ifx\csname url\endcsname\relax
  \def\url#1{\texttt{#1}}\fi
\expandafter\ifx\csname urlprefix\endcsname\relax\def\urlprefix{URL }\fi
\providecommand{\eprint}[2][]{\url{#2}}

\bibitem[{{Almgren} et~al.(2010){Almgren}, {Beckner}, {Bell}, {Day}, {Howell},
  {Joggerst}, {Lijewski}, {Nonaka}, {Singer}, \& {Zingale}}]{ann2010}
{Almgren}, A.~S., {Beckner}, V.~E., {Bell}, J.~B., {Day}, M.~S., {Howell},
  L.~H., {Joggerst}, C.~C., {Lijewski}, M.~J., {Nonaka}, A., {Singer}, M., \&
  {Zingale}, M. 2010, ApJ, 715, 1221.  

\bibitem[{{Barkat} et~al.(1967){Barkat}, {Rakavy}, \& {Sack}}]{barkat1967}
{Barkat}, Z., {Rakavy}, G., \& {Sack}, N. 1967, Physical Review Letters, 18,
  379.

\bibitem[{{Chen}(2014)}]{chen2014r}  
{Chen}, K.-J. 2014, International Journal of Modern Physics D, 23, 1430008. 

\bibitem[{{Chen} et~al.(2013){Chen}, {Heger}, \& {Almgren}}]{chen2013}
{Chen}, K.-J., {Heger}, A., \& {Almgren}, A.~S. 2013, Astronomy and Computing,
  3, 70.

\bibitem[{{Chen} et~al.(2014{\natexlab{a}}){Chen}, {Heger}, {Woosley},
  {Almgren}, \& {Whalen}}]{chen2014psn}
{Chen}, K.-J., {Heger}, A., {Woosley}, S., {Almgren}, A., \& {Whalen}, D.~J.
  2014{\natexlab{a}}, \apj, 792, 44. 

\bibitem[{{Chen} et~al.(2014{\natexlab{b}}){Chen}, {Woosley}, {Heger},
  {Almgren}, \& {Whalen}}]{chen2014ppsn}
{Chen}, K.-J., {Woosley}, S., {Heger}, A., {Almgren}, A., \& {Whalen}, D.~J.
  2014{\natexlab{b}}, \apj, 792, 28. 

\bibitem[{{Cooke} et~al.(2012){Cooke}, {Sullivan}, {Gal-Yam}, {Barton},
  {Carlberg}, {Ryan-Weber}, {Horst}, {Omori}, \& {D{\'{\i}}az}}]{cooke12}
{Cooke}, J., {Sullivan}, M., {Gal-Yam}, A., {Barton}, E.~J., {Carlberg}, R.~G.,
  {Ryan-Weber}, E.~V., {Horst}, C., {Omori}, Y., \& {D{\'{\i}}az}, C.~G. 2012,
  \nat, 491, 228. 
\bibitem[{{Gal-Yam} et~al.(2009){Gal-Yam}, {Mazzali}, {Ofek}, {Nugent},
  {Kulkarni}, {Kasliwal}, {Quimby}, {Filippenko}, {Cenko}, {Chornock},
  {Waldman}, {Kasen}, {Sullivan}, {Beshore}, {Drake}, {Thomas}, {Bloom},
  {Poznanski}, {Miller}, {Foley}, {Silverman}, {Arcavi}, {Ellis}, \&
  {Deng}}]{galyam2009}
{Gal-Yam}, A., {Mazzali}, P., {Ofek}, E.~O., {Nugent}, P.~E., {Kulkarni},
  S.~R., {Kasliwal}, M.~M., {Quimby}, R.~M., {Filippenko}, A.~V., {Cenko},
  S.~B., {Chornock}, R., {Waldman}, R., {Kasen}, D., {Sullivan}, M., {Beshore},
  E.~C., {Drake}, A.~J., {Thomas}, R.~C., {Bloom}, J.~S., {Poznanski}, D.,
  {Miller}, A.~A., {Foley}, R.~J., {Silverman}, J.~M., {Arcavi}, I., {Ellis},
  R.~S., \& {Deng}, J. 2009, \nat, 462, 624. 

\bibitem[{{Heger} \& {Woosley}(2002)}]{heger2002}
{Heger}, A., \& {Woosley}, S.~E. 2002, \apj, 567, 532.
  

\bibitem[{{Hirschi} et~al.(2004){Hirschi}, {Meynet}, \& {Maeder}}]{hirschi2004}
{Hirschi}, R., {Meynet}, G., \& {Maeder}, A. 2004, \aap, 425, 649.
 

\bibitem[{{Kasen} et~al.(2011){Kasen}, {Woosley}, \& {Heger}}]{kasen2011}
{Kasen}, D., {Woosley}, S.~E., \& {Heger}, A. 2011, \apj, 734, 102.
  

\bibitem[{{Kudritzki}(2002)}]{kudri2002}
{Kudritzki}, R.~P. 2002, \apj, 577, 389. 

\bibitem[{{Pan} et~al.(2012){Pan}, {Kasen}, \& {Loeb}}]{pan12a}
{Pan}, T., {Kasen}, D., \& {Loeb}, A. 2012, \mnras, 422, 2701.
   

\bibitem[{{Timmes}(1999)}]{timmes1999}
{Timmes}, F.~X. 1999, ApJS, 124, 241.

\bibitem[{{Timmes} \& {Swesty}(2000)}]{timmes2000}
{Timmes}, F.~X., \& {Swesty}, F.~D. 2000, Ap JS, 126, 501.

\bibitem[{{Weaver} et~al.(1978){Weaver}, {Zimmerman}, \& {Woosley}}]{kepler}
{Weaver}, T.~A., {Zimmerman}, G.~B., \& {Woosley}, S.~E. 1978, ApJ, 225, 1021.

\bibitem[{{Whalen} et~al.(2013{\natexlab{a}}){Whalen}, {Even}, {Frey}, {Smidt},
  {Johnson}, {Lovekin}, {Fryer}, {Stiavelli}, {Holz}, {Heger}, {Woosley}, \&
  {Hungerford}}]{wet12b}
{Whalen}, D.~J., {Even}, W., {Frey}, L.~H., {Smidt}, J., {Johnson}, J.~L.,
  {Lovekin}, C.~C., {Fryer}, C.~L., {Stiavelli}, M., {Holz}, D.~E., {Heger},
  A., {Woosley}, S.~E., \& {Hungerford}, A.~L. 2013{\natexlab{a}}, \apj, 777,
  110. 

\bibitem[{{Whalen} et~al.(2013{\natexlab{b}}){Whalen}, {Even}, {Smidt},
  {Heger}, {Hirschi}, {Yusof}, {Stiavelli}, {Fryer}, {Chen}, \&
  {Joggerst}}]{whalen2013}
{Whalen}, D.~J., {Even}, W., {Smidt}, J., {Heger}, A., {Hirschi}, R., {Yusof},
  N., {Stiavelli}, M., {Fryer}, C.~L., {Chen}, K.-J., \& {Joggerst}, C.~C.
  2014{\natexlab{b}}, \apj, 797, 9.  

\bibitem[{{Whalen} et~al.(2013{\natexlab{c}}){Whalen}, {Fryer}, {Holz},
  {Heger}, {Woosley}, {Stiavelli}, {Even}, \& {Frey}}]{wet12a}
{Whalen}, D.~J., {Fryer}, C.~L., {Holz}, D.~E., {Heger}, A., {Woosley}, S.~E.,
  {Stiavelli}, M., {Even}, W., \& {Frey}, L.~H. 2013{\natexlab{c}}, \apjl, 762,
  L6. 

\bibitem[{{Zhang} et~al.(2011){Zhang}, {Howell}, {Almgren}, {Burrows}, \&
  {Bell}}]{zhang2011}
{Zhang}, W., {Howell}, L., {Almgren}, A., {Burrows}, A., \& {Bell}, J. 2011,
  ApJS, 196, 20. 

\end{thebibliography}

\end{document}